\documentstyle[prd,aps,graphicx]{revtex}
\begin{document}
%*******************************
% 
%   SIMPLE TEXT TRUNCATIONS
%
\def \BE {\begin{equation}}
\def \EE {\end{equation}}
\def \BEAH {\begin{eqnarray*}}
\def \EEAH {\end{eqnarray*}}
\def \BEA {\begin{eqnarray}}
\def \EEA {\end{eqnarray}}
\def \BDM {\begin{displaymath}}
\def \EDM {\end{displaymath}}
\def \t {\tau}
\def \s {\sigma}
\def \bfi {\bar \varphi}
\def \br {\bar \rho}
\def \bz {\bar z}
\def \bt {\bar t}
\def \e {\varepsilon}
\def \g {\gamma}
%*******************************
\draft
\title{Spinning C-metric: radiative spacetime with accelerating, rotating
    black holes}
\author{J. Bi\v c\'ak \footnote{Email address: {\tt bicak@mbox.troja.mff.cuni.cz} }}
\address{Department of Theoretical Physics,
         Faculty of Mathematics and Physics,
         Charles University, \\
         V Hole\v sovi\v ck\'ach  2,
         180 00 Prague 8,
         Czech Republic}
\author{V. Pravda  \footnote{Email address: {\tt pravda@math.cas.cz} }}
\address{Mathematical Institute, 
Academy of Sciences, \protect\\
\v Zitn\' a 25,
115 67 Prague 1, Czech Republic }
\maketitle
\begin{abstract}
The spinning C-metric was discovered by Pleba\'nski and Demia\'nski as a generalization
of the standard C-metric which is known to represent uniformly accelerated  non-rotating
black holes. We first transform the spinning C-metric into Weyl coordinates and analyze
some of its properties as Killing vectors and curvature invariants. A transformation is then
found which brings the metric into the canonical form of the radiative spacetimes with the
boost-rotation symmetry. By analytically continuing  the metric across ``acceleration
horizons", two new regions of the spacetime arise in which both Killing vectors are
spacelike.

We show that this metric can represent two uniformly accelerated, spinning black 
holes, either connected by a conical singularity, or with conical singularities extending
from each of them to infinity. The radiative character of the metric is briefly discussed.
\end{abstract}
\section{Introduction}

The static part of a spacetime representing the standard, non-spinning vacuum C-metric was originally
found by Levi-Civita in 1917-1919  (see references in \cite{Kramer}) but it was only in 1970
when Kinnersley and Walker \cite{KW} understood, by choosing  a better parameterization,
that it can be extended so that it represents two black holes uniformly accelerated
in opposite directions. The ``cause" of the acceleration is given by nodal (conical) singularities
(``strings" or ``struts") along the axis of symmetry. By adding an external gravitational 
field the nodal singularities can be removed \cite{Ernst1}. If the holes are electrically (magnetically) charged
the field causing the acceleration  can be electric  (magnetic) \cite{Ernst2}.  These types of
generalized C-metric have been recently used in the context of quantum gravity - to describe
production of black hole pairs  in strong background fields (see, for example, \cite{Hawking1}, 
\cite{Hawking2}, \cite{Mann}).

From the 1980's  several other works analyzed the standard C-metric. Ashtekar and Dray 
\cite{AshD} were the first to show that the C-metric admits a conformal completion
such that its null  infinity admits spherical sections. Bonnor \cite{Bonnor}  transformed
the vacuum C-metric to a Weyl form, in which the metric represents a ``spherical particle"
(in fact the horizon of the black hole) and a semi-infinite line mass, with a strut  holding
them apart. By another transformation Bonnor enlarged the spacetime so that it became
``dynamic", representing  two black holes (``spherical particles"  in terminology of
\cite{Bonnor}) uniformly accelerated by a spring joining them.  In 1995 Cornish and
Uttley  \cite{Cornish1} simplified Bonnor's procedure  and extended it  to the charged
case \cite{Cornish2}.  Most recently the black-hole uniqueness theorem for 
the C-metric has been proved \cite{Wells}. 

In none of the above works, however, has  the basic fact   been emphasized that the 
vacuum C-metric is just one specific  example  in a large class of asymptotically
flat radiative spacetimes with boost-rotation symmetry (with the boost along the axis of 
rotational symmetry).  
From a unified point of view,
boost-rotation symmetric spacetimes with hypersurface orthogonal axial and
boost Killing vectors were studied geometrically by   Bi\v c\' ak and  Schmidt \cite{BicSch}.
We refer to this detailed work for rigorous definitions and theorems. In fact it is no surprise
that the C-metric was for a long time analyzed in coordinate systems  unsuitable for
treating global issues such as the properties of null infinity. It is algebraically
special, and the coordinate systems were adapted to its degenerate character.
In  polar coordinates $\{t,\ \rho,\ \phi,\ z\}$ the metric  of a general boost-rotation
symmetric spacetime with hypersurface orthogonal axial and boost Killing vectors,
\BE
\frac{\partial}{\partial \phi} \quad {\rm and} \quad z \frac{\partial}{\partial t} +
t \frac{\partial}{\partial z}  \label{Killvec} \ ,
\EE
has the form  (see \cite{BicSch}, Eq. (3.38)):
\BE
{\rm d}s^2 = e^{\lambda} {\rm d} \rho^2 + \rho^2 e^{-\mu} {\rm d} {\phi}^2+
 \frac{1}{z^2-t^2} \left( (e^{\lambda} z^2- e^{\mu} t^2)      {\rm d}z^2 -2zt(e^{\lambda}-e^{\mu}){\rm d}z
{\rm d}t - (e^{\mu} z^2 -e^{\lambda} t^2 ){\rm d}t^2  \right) \ , \label{bsrotmA}
\EE
where $\mu$ and $\lambda$ are functions of $\rho^2$ and $z^2-t^2$, satisfying, as a 
consequence of Einstein's vacuum equations, a simple system of three equations, one
of them being the wave equation for the function $\mu$. 
The whole structure of the group orbits in boost-rotation symmetric curved spacetimes  outside
the sources (or singularities) is the same as the structure of the orbits generated by the
axial  and boost Killing vectors in Minkowski space.  In particular, the boost Killing
vector  (\ref{Killvec})  is timelike  in the region $z^2>t^2$.  It is this region which can
be transformed  to the static Weyl form.  Physically this corresponds to the transformation
to ``uniformly accelerated frames" in which sources are at rest and the fields are time independent. 

However, in the other ``half" of the spacetime, $t^2>z^2$  (``above  the roof" in the terminology of 
\cite{BicSch}),  the boost Killing vector is spacelike so that in this region the metric 
(\ref{bsrotmA})  is nonstationary. It can be shown  that for $t^2>\rho^2+z^2$  the metric can be
locally  transformed into the metric  of the Einstein-Rosen waves. The  radiative properties
of the specific boost-rotation symmetric spacetime were investigated long  ago
\cite{Bic68}, and for more general metrics in  \cite{BicT}, \cite{BicB}, \cite{PP}. The 
boost-rotation symmetric spacetimes  were used as tests beds
in numerical relativity - see the review \cite{LesHouches} for more details and references.

Now in all the work mentioned above it was assumed that the axial and boost Killing vectors
are hypersurface orthogonal. Recently,  Bi\v c\'ak and    Pravdov\'a
\cite{BicPra}  analyzed symmetries compatible with asymptotic flatness and
admitting gravitational and electromagnetic radiation.  They have shown that in 
axially symmetric electrovacuum spacetimes in which at least locally a smooth null infinity
exists, the only second allowable  symmetry which admits  radiation  is the boost symmetry.
The axial and an additional Killing vector have {\it not}  been  assumed to be hypersurface
orthogonal.  In \cite{BicPra}  the general functional forms of gravitational  and electromagnetic
news functions, and of the total mass of asymptotically flat boost-rotation symmetric spacetimes
at null infinity have been obtained. However, until now no general theory  similar to that
given in \cite{BicSch}  in the hypersurface orthogonal case is available for the boost-rotation
symmetric  spacetimes with Killing vectors which are not hypersurface orthogonal.
 Nevertheless
there is one explicitly given metric which can be expected to serve as an example 
of these spacetimes - the spinning vacuum C-metric.
It was discovered by Pleba\'nski and Demia\'nski \cite{Pleb}  in 1976, studied later by
Farhoosh and Zimmerman  \cite{FZ}, and very recently briefly discussed by Letelier
and Oliveira \cite{Letelier}. In none of these works, however,  was the boost-rotation
character  of  the spacetime properly  analyzed.  In particular, ``the canonical
coordinates"  in which the metric   represents
a generalization of the metric (\ref{bsrotmA}) so that global issues outside the
sources could properly 
be studied have not been found so far.

The main purpose  of this paper is to find  such  a  representation of the spinning 
C-metric (SC-metric)  which  would  generalize (\ref{bsrotmA}) and, thus,  could
also serve as a convenient example  for building up the general theory of
boost-rotation symmetric spacetimes with the Killing vectors which are in general
not hypersurface orthogonal.

In the next section we first write  the Pleba\'nski-Demia\'nski class of metrics
\cite{Pleb},  specialized to the  spinning vacuum case, discuss the ranges of parameters
entering  the metric and indicate the limiting
procedure leading to the Kerr metric.  In Section III the SC-metric is transformed into Weyl
coordinates.
This is not an easy task, fortunately we succeeded in  generalizing the  procedure Bonnor
\cite{Bonnor} used for the standard C-metric without spin. We show that,
by choosing  different values of the original  Pleba\'nski-Demia\'nski coordinates we can,
in principle, arrive at various  Weyl spacetimes.

The properties of Killing vectors and of some invariants of the Riemann tensor
lead us, in section IV,  to choose a physically plausible Weyl spacetime
which contains  both the black-hole  and the acceleration horizon. Similarly as has been done
with  the standard  C-metric  \cite{Bonnor}, \cite{Cornish1}, we concentrate on this Weyl
portion.
In section V the metric  is transformed into the ``canonical form"  of boost-rotation
symmetric spacetimes with Killing vectors which need not be hypersurface orthogonal
(see Eqs. (\ref{BStvar})-(\ref{trel})).
By analytically  continuing the resulting metric across the acceleration horizons
(``the roof"),  two new regions of spacetime arise as in the standard case 
of the C-metric without spin.

Let us emphasize that it is only now, after having described the spacetime  in coordinates
in which global issues can be addressed,  that we are able to fix coordinates $t$, $\phi$
and two Killing vectors appropriately. The importance of this fact is briefly discussed
in relation to the most recent work \cite{Letelier}  in which a transformation to the
canonical form was not performed. We also show that, analogous to the C-metric 
without spin, the axis of symmetry contains nodal singularities
between the spinning ``sources" (holes)  which cause  the acceleration, and is regular  elsewhere;
or the axis can be
made regular between the sources but then the nodal  singularities  extend from 
each of the sources to infinity (see Fig. 4  a, b).

Finally, in our concluding remarks we briefly compare our results with those obtained recently
in \cite{Letelier}. We then give  two figures, constructed numerically, 
exhibiting clearly the radiative character of the SC-metric. A detailed  analysis 
of the radiative properties  of the SC-metric  and of its analytic extension  into 
the holes' interiors will be given elsewhere.

\section{Spinning C-metric}
 Pleba\'nski and Demia\'nski \cite{Pleb} gave their class of metrics  in the coordinates $\{\tau$, $p$, $q$, $\sigma\}$ in the form
\BEA
{{\rm d}s}^{2}={\frac {E\left (Q-P{q}^{4}\right )}
{F}} {{ d\t}}^{2}-{\frac {EF}{Q}} {{\rm d}q}^{2}-{\frac {EF}{P}}
{{\rm d}p}^{2}  %+\nonumber \\
+{\frac {E\left (Q{p}^{4}-P\right )}{F}} {{\rm d}\s}^{2}-{\frac {2\,
E\left (Q{p}^{2}+P{q}^{2}\right )}{F}} {\rm d}\t {\rm d}\s ,
\label{RotC}
\EEA
where
\BEA
P&=&\g-{ \e}\,{p}^{2}+2\,m{p}^{3}-\g{p}^{4} \label{eqP} \ , \\
Q&=&-\g+{ \e}\,{q}^{2}+2\,m{q}^{3}+\g{q}^{4} \label{eqQ} \ ,  \\
E&=&\left (p+q\right )^{-2} \ , \label{eqE} \\
F&=&1+{p}^{2}{q}^{2} \ , \label{eqF}
\EEA
$m$, $\g$, $\e$ being constants. The choice of the exact ranges of dimensionless coordinates
$p$, $q$ will be specified later, at this point we consider $p$, $q$ $\in$ $R$; the coordinates
$\tau$, $\sigma$ have dimension of $($length$)^2$, and we also take $\tau$, $\sigma$ $\in R$.
In the  general form of Pleba\'nski-Demia\'nski metric (Eqs. (2.1), (3.25) in \cite{Pleb}) we put the parameters
$n$, $e$, $g$, $\lambda$, corresponding to the NUT parameter, electric and magnetic charge, and 
 cosmological constant, equal to zero. If parameter $m=0$ the spacetime is flat. It is not straightforward
to obtain standard metrics from (\ref{RotC}). In \cite{Pleb} it is shown (see also \cite{Kramer}) that
the stationary (non-accelerated) Kerr solution can be obtained by scaling  the coordinates
and, simultaneously, the parameters as
\BDM
p=\frac{p'}{l} \ , \ \  q=\frac{-l}{q'} \ , \ \  \tau={l}{\tau'} \ , \ \  \sigma={l^3}{\sigma'} \ , 
\EDM
\BE
m={l^{-3}}{m'} \ , \ \  \e={l^{-2}}{\e'} \ , \ \  \g={l^{-4}}{\g'} \ .
\EE
Then, after taking the limit $l \rightarrow  \infty$ and putting
\BE
m'=M \ , \ \  \e'=1 \ , \ \  \g'=a^2 \ , \ \ p'=-a \cos \theta \ , \ \ q'=r \ , \ \ \sigma' = -\phi/a \ , \ \  \tau' =
t + a \phi \ ,
\EE
we obtain the standard Kerr metric  with mass $M$ and specific angular momentum $a$ in 
the Boyer-Lindquist coordinates. 
 If $\e = m^2/\g$, the limiting procedure leads to the extreme
Kerr hole with $a^2=M^2$.
It is worth  noticing  that in this case   the quartic
$P$ can be factorized, $ P=-{\g}^{-1}(\g p^2-mp+\g)(\g p^2 -m p - \g) $.
 If $\e > m^2/\g$, we obtain a general Kerr black hole; with  $\e < m^2/\g$
- a Kerr naked singularity. 
Another limiting procedure (see \cite{Pleb}, \cite{Kramer})
leads to the C-metric without rotation.

In the following we shall restrict ourselves  to the general case  in which the polynomials (\ref{eqP})
and (\ref{eqQ}) have four  different real roots.  We thus omit the special  cases 
which may be of interest (cf. \cite{AshD}, \cite{Bonnor2}) 
but will be dealt with elsewhere.  Notice that if $p_i$ is the root of the polynomial  $P$ given
by Eq. (\ref{eqP}),  then $q_i=-p_i$ is the root of the polynomial $Q$   (Eq. (\ref{eqQ})).
The analysis of the 4th-order polynomial (\ref{eqQ}) shows that if $\g$ is chosen positive, the polynomial
has four different roots provided that
\BEA
 \epsilon &>& 2 \sqrt{3} \gamma \nonumber \ , \\
 \frac{\epsilon (\epsilon^2+36 \gamma^2) -(\epsilon^2-12 \gamma^2)^{3/2}}{54 \gamma}  < &m^2& 
< \frac{\epsilon (\epsilon^2+36 \gamma^2) + (\epsilon^2-12 \gamma^2)^{3/2}}{54 \gamma} \ .
\EEA 
In Fig. 1 we illustrate the allowed range of the parameters for $\g=1$. 
\begin{figure}
\begin{center}
\includegraphics*[height=2in]{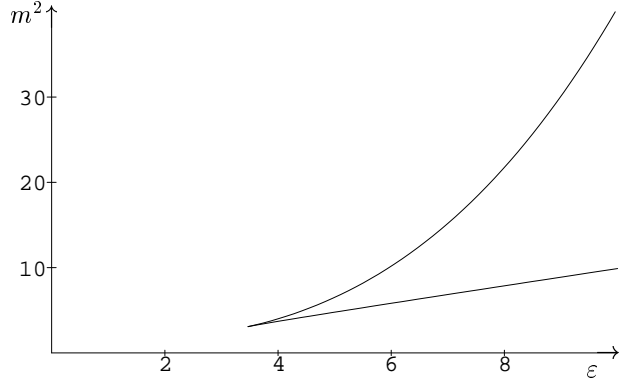}
%\put(-165,33){$\varepsilon$}
%\put(-288,120){$m^2$}
\end{center}
\caption{The allowed range of  parameters $\e$  and $m^2$ for $\g=1$. If  a point $(\e,m^2)$ is ``between" the  two curves, then both polynomials $P, Q$ have four distinct real roots.}
\end{figure}

Returning back  to the metric (\ref{RotC}) - (\ref{eqF}), we immediately observe
that it has two Killing vectors,
\BE
{\frac{\partial}{\partial \tau}}  \quad {\rm and} \quad  {\frac{\partial}{\partial \sigma}} \ . \label{Killpq}
\EE

\section{SC-metric in Weyl coordinates}

We just saw that the SC-metric has two Killing vectors (\ref{Killpq}). We also noticed that in  a limit
it goes over into  the C-metric without rotation (cf. \cite{Kramer}) which is known to be axisymmetric and in some 
regions static. Hence, we expect that the SC-metric  can be converted into a standard Weyl-Papapetrou
form
\BE
{\rm d}s^2 ={\rm e}^{-2U} \left ({{\rm e}^{2 \nu}} ({{\rm d} \br}^{2}
+{{ {\rm d} \bz}}^{2} )+{\br}^{2}{{{\rm d}\bfi}}^{2} \right )-{{\rm e}^{2\,U}}
\left ({ {\rm d} \bt}+A{ {\rm d} \bfi}\right )^{2}  \ ,   
\label{Weyl}
\EE
where  functions $U$,  $\nu$, $A$ depend only on
$\br$ and $\bz$, $\bfi \in \langle 0, 2 \pi)$,  $\br \in \langle 0,\infty)$, $\bt,\bz \in R$.

Vacuum Einstein's equations imply
\BEA
%%%%%%%%%%%%%%%%%%%%%%%%%%FIRST
\frac{\partial^2 U}{\partial \bz^2} + \frac{1}{\br} \frac{\partial U}{\partial \br}
+ \frac{\partial^2 U}{\partial \br^2} = -\frac{1}{2} \frac{{\rm e}^{4U}}{\br^2}
\left [ \left (\frac{\partial A}{\partial \br} \right )^2 +
\left (\frac{\partial A}{\partial \bz}  \right )^2 \right ] \ , \nonumber \\
%%%%%%%%%%%%%%%%%%%%%%%%%%SECOND
\frac{\partial }{\partial \bz} \left ( \frac{{\rm e}^{4U}}{\br}
\frac{\partial A}{\partial \bz} \right ) +
\frac{\partial }{\partial \br} \left ( \frac{{\rm e}^{4U}}{\br}
\frac{\partial A}{\partial \br} \right ) = 0 \ , \nonumber \\
%*************************THIRD
\frac{1}{\br} \frac{\partial \nu}{\partial \br} =
\left (\frac{\partial U}{\partial \br} \right )^2
-\left (\frac{\partial U}{\partial \bz} \right )^2 -\frac{{\rm e}^{4U}}{4 \br^2}
\left [ \left (\frac{\partial A}{\partial \br} \right )^2 -
\left (\frac{\partial A}{\partial \bz}  \right )^2 \right ] \ , \\
%************************FOURTH
\frac{1}{\br} \frac{\partial \nu}{\partial \bz} = 2 \frac{\partial U}{\partial \br}
\frac{\partial U}{\partial \bz} - \frac{{\rm e}^{4U}}{2 \br^2}
\frac{\partial A}{\partial \br} \frac{\partial A}{\partial \bz} \ . \nonumber
\EEA
Axial and timelike Killing vectors
\BE
\xi_{(\bfi)} = \frac{\partial}{\partial \bfi} \ , \   \xi_{(\bt)} = \frac{\partial}{\partial \bt}  \ , \label{KillWeyl}
\EE
determine invariantly the cylindrical-type radial coordinate by the relation
\BE
{\br}^{2}={ g_{\bt \bfi}}^{2}-{ g_{\bt \bt}}\,{ g_{\bfi \bfi}} \ = \ (\xi_{(\bt)} , \xi_{(\bfi)} )^2 -
(\xi_{(\bt)} , \xi_{(\bt)} ) (\xi_{(\bfi)} , \xi_{(\bfi)} ) \ .
\label{Weylrho}
\EE

In the following we shall consider points with $\br = 0$ as ``lying on the $\bz$-axis of axial
symmetry". This is indeed the axis of Weyl's coordinates, however, it must be born on mind
that  $\br = 0$ is a real smooth geometrical (physical) axis only there where the metric (\ref{Weyl})
satisfies the regularity conditions
\BE
\lim_{\br \to \infty} \frac{X_{,\alpha} X^{,\alpha}}{4X} = 1\ , \  {\rm where}\  X =\xi_{(\bfi) \alpha} \xi_{(\bfi) }^\alpha  \ , \  \ \alpha=0-3.   \label{regosa}
\EE
We shall see  that for some intervals of $\bz$ the points  $\br=0$ may represent
either a horizon of a rotating black hole or a rotating string - regularity conditions
(\ref{regosa}) are then not satisfied. The norms of the Killing vectors will enable us to
distinguish between horizons and strings.

In order to convert the original metric  (\ref{RotC})-(\ref{eqF}) into the Weyl-Papapetrou
metric (\ref{Weyl}),
we shall  seek the transformation of the form
\BE
\br = \br(p,q) \ , \ \ \bz = \bz(p,q) \ , \ \ \bt = \bt(\tau, \sigma) \  ,  \ \  \bfi = \bfi (\tau, \sigma) \  . 
\label{transform}
\EE
Since the  Killing vectors (\ref{Killpq}) must be linear combinations with constant coefficients
of the Killing vectors (\ref{KillWeyl}),  we can write
\BEA
{ d\t}={ \kappa_1}\,{ d\bt}+{ \kappa_2}\,{ d\bfi} \ , \nonumber \\
{ d\s}={ \kappa_3}\,{ d\bt}+{ \kappa_4}\,{ d\bfi} \ , \label{eqtrKill}
\EEA
with  $ \kappa_1 \dots  \kappa_4$ constants. As a consequence of  (\ref{transform}) and
(\ref{eqtrKill}) we find  the metric components associated with the Killing vectors to be
given  in Weyl 
coordinates  by
\BEA
g_{\bfi \bfi}&=&\frac{E}{F} \left( \left(-\kappa_2+p^2 \kappa_4   \right)^2 Q-
\left(\kappa_4+q^2 \kappa_2  \right)^2 P \right)  \ , \nonumber \\
g_{\bfi \bt}&=&  \frac{E}{F}
\left (\left (-{\kappa_1}+{p}^{2}{ \kappa_3}\right )\left (-{
\kappa_2}+{p}^{2}{\kappa_4}\right )Q-\left ({ \kappa_3}+{q}^{2}{ \kappa_1}\right )
\left ({ \kappa_4}+{q}^{2}{\ \kappa_2}\right )P\right )   \ , \nonumber \\
g_{\bt \bt}&=&\frac{E}{F} \left (\left (-{ \kappa_1}+{p}^{2}{ \kappa_3}\right )^{2}Q-\left ({
 \kappa_3}+{q}^{2}{ \kappa_1}\right )^{2} P\right )  \ . \label{gCcompts}
\EEA
Substituting these components into Eq. (\ref{Weylrho}), we arrive at surprisingly simple expression
for $\br$:
\BE
\br^2={E}^{2}{\cal K}^{2}PQ \ ,     
\ \ \  {\cal K}  =\left ({ \kappa_2}\,{\kappa_3}-{ \kappa_1}\,{ \kappa_4}\right ) \ . 
\label{rhopq}
\EE
Notice that this  result  is in a disagreement with the last Eq. (34) of \cite{Letelier}; there
only the first term should appear. 

Turning now to the transformation of variables $p$, $q$ into Weyl's $\br$, $\bz$,  we first write 
\BEA
{ d\br}={\cal A}{ dp}+{\cal B}{ dq} \ , \nonumber \\
{ d\bz}={\cal C}{ dp}+{\cal D}{ dq}  \ .   \label{transf2}
\EEA
Functions ${\cal A}$ and ${\cal B}$ can be determined from Eq. (\ref{rhopq}).
Functions ${\cal C}$ and ${\cal D}$ can be obtained by inverting
relations (\ref{transf2}), substituing into the original metric (\ref{RotC}) and demanding that
$g_{\br \br} =  g_{\bz \bz}$, $ g_{\br \bz} = 0 $.
In this way we  find 
\BEA
{\cal A}={\cal K} \frac{\partial}{\partial p} \left( E {\sqrt{P Q}}\right) \ &,& \quad
{\cal B} ={\cal K} \frac{\partial}{\partial q} \left( E {\sqrt{P Q}}\right) \ , \\ \nonumber
{\cal C}=\mp {\frac {{\cal B}\sqrt {QP}}{P}} \ &,& \quad
{\cal D}=\pm {\frac {{\cal A}\sqrt {QP}}{Q}} \ .  
\EEA
Since ${\cal C} = \partial \bar z /  \partial p$,  ${\cal D} = \partial \bar z /  \partial q$, 
and the integrability conditions $\partial {\cal C} / \partial q - \partial {\cal D} / \partial p = 0$
are satisfied, we can integrate for $\bz$. Regarding Eqs. (\ref{eqP}) - (\ref{eqF}) we obtain 
\BE
\bz= \mp |{\cal K}|\left( {\frac{-\g-{ \e}\,qp-m{q}^{2}p+mq{p}^{2}+
\g{q}^{2} {p}^{2} }{\left (p+q\right )^{2}}} \right ) \ ,
\label{vysltrans}
\EE
where we put an additive constant equal to zero. We thus arrive at the metric functions
determining   the Weyl-Papapetrou line element (\ref{Weyl}) in the form:
\BEA
{\rm e}^{2U} &=& -g_{\bt \bt} \label{e2Upq} \ ,  \\
{\rm e}^{2\nu} &=& -g_{\br \br}  g_{\bt \bt} \ , \label{e2Npq} \\
A &=& \frac{g_{\bt \bfi}}{g_{\bt \bt}} \ ,  \label{apq}
\EEA
where
\BE
g_{\br \br} = g_{\bz \bz} = -\frac{E F}{A^2 P + B^2 Q} \ , \label{grrpq}
\EE
and $g_{\bt \bt}$, $g_{\bt \bfi}$ are given by Eqs. (\ref{gCcompts}). In this way
we succeeded in expressing the Weyl metric functions in terms of $p$, $q$.

In order to find these functions directly in the Weyl  coordinates, we have to invert
relations (see Eqs. (\ref{rhopq}), (\ref{vysltrans}) and (\ref{eqP}), (\ref{eqQ}))
\BEA
\bz &=& {\cal K} {\frac { -\gamma-\varepsilon\,qp-m{q}^{2}p+mq{p}^{2}+\gamma{q}^{2}{p}^{2}
 }{ (p+q )^{2}}} \ , \label{tranz} \\
{\br}^{2} &=& {\frac {{\cal K}^{2}(\g-{ \e}\,{p}^{2}+2\,m{p}^{3}-\g{p}^{4})(-\g+{ \e}\,{q}^{2}+2\,m{q}^{3}+\g{q}^{4})}{(p+q)^{4}}} \label{tranr} \ .
\EEA
This is not an easy task. Fortunately, we can try to generalize the procedure used by Bonnor 
\cite{Bonnor} for the vacuum C-metric without rotation  (see Eq. (11) in \cite{Bonnor}).
We thus wish to find constants $\alpha_1$, $\alpha_2$, $\alpha_3$, $\alpha_4$ and $z_i$, $i=1,\ 2, \ 3$,
such that 
\BE
\br^2 + (\bz-{\bz_i})^2 = \Bigl(\frac{{\alpha_1}p+{\alpha_2}q+{\alpha_3}+{\alpha_4} pq}{p+q}\Bigr)^2  \ .
\label{pozadavek}
\EE
Multiplying the last equation by $(p+q)^2$, regarding expressions (\ref{tranz}), (\ref{tranr}), and requiring the coefficients
in the resulting polynomial  to vanish, we get
\BE
{\alpha_1} =-W_i \ , \quad
{\alpha_2} =W_i  \ , \quad
{\alpha_3} =\frac{{\cal K}^2 m \gamma}{W_i} \ ,  \quad
{\alpha_4} =\frac{{\cal K} m \bz_i}{W_i}  \ , \label{alphas} 
\EE
where $\bz_i/{\cal K}$ are the roots of the equation
\BE
2(\bz_i/{\cal K})^3+\varepsilon (\bz_i/{\cal K})^2+ 2\gamma^2 (\bz_i/{\cal K}) + \varepsilon \gamma^2 -m^2 \gamma = 0 \ , \ \ i=1,\ 2,\ 3,  \label{cubeq}
\EE
and
\BE
W_i =\sqrt{{\cal K}^2{\gamma}^{2}+{\bz_i}^2}   \ .
\EE
Next we introduce functions
\BE
R_i  =  {\sqrt{\br^2 + (\bz-{\bz_i})^2 }   \  \geq \ 0 \ . }  \label{Rirz}
\EE
After substituting for $\alpha_1 \dots \alpha_4$ from Eq. (\ref{alphas})  into the relation (\ref{pozadavek}) and taking square
roots, we can rewrite the functions $R_i$ in the form
\BE
R_i  =  \epsilon_i \frac{1}{(p+q)} \left(-W_i p+ W_i q + \frac{{\cal K}^2 m\gamma}{W_i}+
\frac{{\cal K} m \bz_i pq}{W_i} \right) \ , \label{eqRipq}
\EE
in which $ \epsilon_i$ is chosen $+1$ or $-1$ so that the right-hand side of 
(\ref{eqRipq}) is indeed positive. 
Eq. (\ref{eqRipq})  represents three dependent equations for $p$ and $q$ as functions
of $\br$ and $\bz$. These can be easily solved if first simple new variables such as, for example, 
$pq$, $p+q$, $-p+q$ are introduced. Finally we succeed to express $p$ and $q$ as 
follows: 
\BEA
p &=& \frac{S_1-S_2}{2S_3} \ ,   \nonumber \\
q &=& \frac{S_1+S_2}{2S_3} \ , \label{pqrz}
\EEA
where
\BEA
S_1 &=& -\epsilon_1 \epsilon_2 \epsilon_3  {\cal K}^2 m \gamma(\bz_1-\bz_2) (\bz_1-\bz_3) (\bz_2-\bz_3)
\ , \nonumber \\
S_2 &=& \gamma m {\cal K}^2 \left ( \epsilon_2 \epsilon_3(\bz_3-\bz_2)W_1 R_1+\epsilon_1
\epsilon_3 (\bz_1-\bz_3)
W_2 R_2-\epsilon_1  \epsilon_2 (\bz_1-\bz_2) W_3 R_3 \right )   \nonumber \ ,  \\
S_3 &=&  - \epsilon_2 \epsilon_3 (\bz_3-\bz_2) ( {\cal K}^2 \gamma^2 - \bz_2 \bz_3) W_1 R_1
          + \epsilon_1 \epsilon_3 (\bz_3-\bz_1) ( {\cal K}^2 \gamma^2 - \bz_1 \bz_3) W_2 R_2
           \label{Spq} \\
         && + \epsilon_1 \epsilon_2 (\bz_1-\bz_2) ( {\cal K}^2 \gamma^2 - \bz_1 \bz_2) W_3 R_3 
\ .   \nonumber
\EEA
Therefore, by choosing different values for $\epsilon_i$ in the last expressions, we arrive at different
$p$'s and $q$'s according to relations (\ref{pqrz}) and, thus, to different Weyl metric functions
(\ref{e2Upq}) - (\ref{apq}).  Because there exist eight different  choices for 
$\epsilon_i$ we get in principle eight different Weyl spacetimes. Their properties will be
analyzed in the next section. As we shall see, only four of them have the signature $+2$,
which we are  using in this paper.
\section{Properties of the SC-metric in \lowercase{$(p,q)$} and $(\br,\bz)$ coordinates }
First we shall study the character of the Killing vectors. In particular we seek a  region
of spacetime
in which there exists  a linear combination (with constants coefficients $k_1, k_2$) of Killing 
vectors, 
\BE
\eta = k_1 \frac{\partial}{\partial \tau}  + k_2  \frac{\partial}{\partial \sigma} \label{combkill} \ ,
\EE
which is spacelike, and another combination  which is timelike so that the
region is then stationary. Hence,  the quadratic form 
\BE \label{norma}
 g_{\alpha \beta } \eta^{\alpha } \eta^{\beta } = k_1^2 g_{\tau \tau} + 2 k_1 k_2 g_{\tau \sigma }
+ k_2^2 g_{\sigma \sigma } \label{quadform}
\EE
must  there be indefinite. This implies that the matrix $g_{AB}$, $A,\ B=\tau, \sigma,$ 
(given by  Eqs. (\ref{RotC}) - (\ref{eqF})) has eigenvalues
$\lambda_1, \lambda_2$ satisfying  $\lambda_1  \lambda_2 < 0$. Since it turns out that 
\BE
\lambda_1  \lambda_2 = -E^2 P Q \ , \label{prodlambd}
\EE
 the stationary regions are those in which
\BE
PQ>0 \ .\label{statcond}
\EE
The analysis  of the eigenvalues $\lambda_1,  \lambda_2 $ reveals that $ \lambda_1 + \lambda_2 =- (E/F) (P(1+q^4) -Q(1+p^4))$ and since sign($\lambda_1  \lambda_2 $) = $-$sign($PQ$), we find out that  the metric (\ref{RotC})
has the signature $-2$ if $P>0$, whereas $P<0$ implies signature $+2$.
Since in this paper  we are choosing the signature $+2$, we take $P<0$.

The  character of the regions described by the specific intervals of $(p,q)$ coordinates is best understood
by plotting them  in the schematic Figure 2.  Here $p_1 < p_2 < p_3 < p_4$ and $q_1 < q_2 < q_3 < q_4$
denote the roots of the polynomials $P$ and $Q$, the numbers $1,\  \dots, \ 5$ indicate regions
between individual roots or between a root and infinity. Notice that the Figure is schematic in the sense
that the squares are plotted to have the same size, even though the lengths of the intervals between the roots
are not the same.  
\begin{figure}
\begin{center}
\includegraphics*[height=6cm]{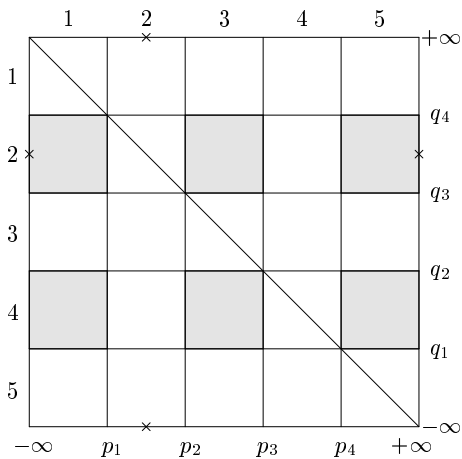}
\end{center}
\caption{The schematic illustration of the character of the regions described by the specific intervals of $(p,q)$ coordinates. See the text for the details.}
\end{figure}

Now in the polynomials $P$, $Q$ we choose, without loss of generality, parameter $\g > 0$.  Since our signature
$+2$ requires $P<0$, we consider only the squares in the columns $1, \ 3, \ 5.$ In the squares which 
represent  stationary regions, the inequality (\ref{statcond})  is valid -  in Fig. 2 these squares
are shaded. By numerically analyzing the right-hand side of Eq. (\ref{eqRipq}) we can make sure that
in the squares $\{2,1\}$ and $\{2,5\}$ all parameters $\epsilon_i=-1$, in $\{4,1\}$ and $\{4,5\}$,
$\epsilon_1= \epsilon_2=1$, $\epsilon_3=-1$, in $\{4,3\}$,  $\epsilon_1= \epsilon_3=1$, $\epsilon_2=-1$
and in the square $\{2,3\}$, which will play the main role in  the following, we find $\epsilon_1=-1$, $\epsilon_2= \epsilon_3=1$.
%************
\begin{figure}
\begin{center}
\includegraphics*[height=2in]{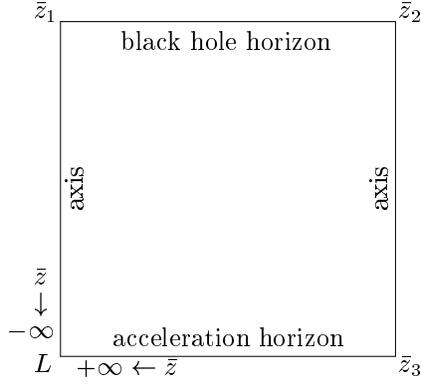}
\end{center}
\caption{The square $\{2,3\}$, with $p \in \langle p_2 , p_3 \rangle$,  $q \in \langle q_3, q_4 \rangle.$
The locations of infinity, the axis, and the black-hole and acceleration horizon
are indicated. See the text for the details.}
\end{figure}

The shaded square $\{2,3\}$ is illustrated in detail in Fig. 3. 
The boundaries of the square represent the axis $\br=0$ because of the relation 
(\ref{rhopq}).  The transformation (\ref{tranz}) implies that the vortices of the square 
correspond to the roots $\bz_i$ of the equation (\ref{cubeq}). From the same transformation
 (\ref{tranz}) we also see that the lower left vortex  $L$ (corresponding to $p_2,\ q_3$) is special:
when it is approached along two different edges of the square we arrive at either
$\bz \rightarrow + \infty$ or $\bz \rightarrow - \infty$. In addition, by approaching this vortex along
different lines from the interior of the squares we can achieve various values of $\bz$ and $\br$.

Since the character of the Killing vectors depends on the eigenvalues $\lambda_1$, $\lambda_2$ 
(cf. Eqs. (\ref{quadform}), (\ref{prodlambd})),  and on the boundaries of the square either $\lambda_1=0$  or $\lambda_2=0$,  just one of the Killing vectors has to be null there. Between $\bz_1$ and $\bz_2$ all other
Killing vectors are spacelike so that this part of the axis will describe the Killing horizon,  the same
being true for the boundary between $L$ and $\bz_3$.  Later we shall see that the segment between
$\bz_1$ and $\bz_2$ corresponds to the black-hole horizon whereas that between $L$ and $\bz_3$
to the acceleration horizon. The remaining segments, between $\bz_1$ and $L$, and $\bz_2$ and $\bz_3$,
describe the axis of symmetry, $\br=0$, along which, however, may lay the string represented by a conical
singularity. There exists just one combination (\ref{combkill}) of the Killing vectors (corresponding
to the minimum of the quadratic form (\ref{quadform}))  the norm of which is null, all other combinations
(\ref{combkill}) give timelike vectors.

It is useful to calculate several simplest invariants of the Riemann tensor for the metric (\ref{RotC}):
\BEA
I_1 &=& {R^{\alpha \beta}}_{\gamma \delta} {R_{\alpha \beta}}^{\gamma \delta}
= -48 m^2 \frac{(q^6 p^6-15q^4p^4+15q^2p^2-1)(p+q)^6}{(1+p^2q^2)^6} \ , \nonumber \\
I_2 &=& {R^{\alpha \beta}}_{\gamma \delta}  {R^{\gamma \delta}}_{\varepsilon \lambda}
{R^{\varepsilon \lambda}}_{\alpha\beta} = 96 m^3 \frac{(9q^8p^8-84q^6p^6+126q^4p^4-36p^2q^2+1)
(p+q)^9}{(1+p^2 q^2)^9} \ ,  \label{invars}\\
I_3 &=& R^{\alpha\beta\gamma\delta}R_{\alpha\varepsilon\lambda\mu}
{R^{\varepsilon\lambda}}_{\beta\nu} {R^{\mu\nu}}_{\gamma\delta} =
-144m^4 \frac{(p+q)^{12}}{(1+p^2q^2)^6} \ . \nonumber
\EEA
Curiously, it is the last invariant which most simply  indicates that asymptotically flat regions can  exist
only on the line given by 
\BE
p+q=0 \ .
\EE
On this line invariants $I_1$ and $I_2$ also vanish. In Section V, where the ``canonical coordinates" will be introduced,
we shall see that indeed the point $L$ in Fig. 3 corresponds to ``points at infinity" where the spacetime is
flat. The invariants (\ref{invars}) diverge at the points 
\BE
  [p,q]=[0,\infty], \
[0,-\infty], \ [-\infty,0], \  [\infty,0],
\EE
which thus correspond to curvature singularities.

The region $\{2,3\}$ described above appears to be most plausible from physical point of view to represent
a uniformly accelerated, rotating black hole.  We shall use this region in the following.  The 
regions  $\{2,1\}$, $\{2,5\}$  which also give signature $+2$ and contain  a timelike Killing
vector,  imply the existence of a naked singularity.
Indeed, the invariants (\ref{invars}) diverge at the point $[p=-\infty,q=0]$ which lies in the left  edge of
$\{2,1\}$, and at the point $[p=\infty,q=0]$  in the right edge of $\{2,5\}$. In fact one can make sure that
by gluing these to edges together (and thus identifying the singular points) both regions $\{2,1\}$ and
$\{2,5\}$ glued together will give one Weyl space.  (Notice that $\epsilon_1$, $\epsilon_2$ and $\epsilon_3$
are the same in these regions.) Since, however, it contains a singular ring at $\br \not= 0$
(corresponding to the identified singular points at the edges) we shall not consider these regions further.
We shall see in the next section that the region  $\{2,3\}$, after appropriate transformation and extension,
gives the required  form of the boost-rotation symmetric  spacetime with Killing vectors which
are not hypersurface orthogonal.

\section{ The SC-metric in the canonical coordinates adapted to the boost-rotation symmetry}

Following Bonnor's analysis \cite{Bonnor}  of the standard C-metric we now make the transformation
from the Weyl  coordinates to new coordinates in which the boost-rotation symmetry of the SC-metric
will become manifest. Simultaneously, by analytically continuing the resulting metric, two new regions
of spacetime will arise - in a close analogy to the C-metric without spin. 
The transformation 
\BEA
\bar t &=& {\rm arctanh} \frac{t}{z} \ , \nonumber \\
\br &=& \rho \sqrt{z^2-t^2} \ ,  \label{transfT1} \\
\bz - \bz_3 &=& \frac{1}{2} \left( \rho^2 + t^2 -z^2 \right) \ , \nonumber  \\
\bfi &=& \varphi \ , \nonumber 
\EEA
brings the metric (\ref{Weyl}) into the form
\BEA
{\rm ds}^2 = {\rm e}^{\lambda} {\rm d} \rho^2 + \rho^2 {\rm e}^{-\mu} {\rm d} \varphi^2 +
\frac{1}{z^2-t^2} \left[ ({\rm e}^{\lambda} z^2 - {\rm e}^{\mu} t^2 ) {\rm d} z^2 -   2zt ({\rm e}^{\lambda} - {\rm e}^{\mu}  ) {\rm d} z {\rm d} t +     ({\rm e}^{\lambda} t^2 - {\rm e}^{\mu} z^2 ) {\rm d} t^2  \right] \nonumber \\ 
-    2A {\rm e}^{\mu} (z {\rm d}t -  t {\rm d} z)  {\rm d} \varphi -A^2 {\rm e}^{\mu} (z^2-t^2)   {\rm d} \varphi^2 \ ,
\label{BStvar}
\EEA
where 
\BE
{\rm e}^{\mu} = \frac{{\rm e}^{2U}}{z^2-t^2}  \  , \quad
{\rm e}^{\lambda} = 
\frac{{\rm e}^{2 \nu}}{{\rm e}^{2 U}} (\rho^2 + z^2 - t^2) \ ,  \label{trel}
\EE
and functions  ${\rm e}^{2U}$, ${\rm e}^{2 \nu}$ and $A$ are given in terms of coordinates 
$\{ t, \rho, z, \varphi \}$ by Eqs. (\ref{gCcompts}), 
(\ref{e2Upq})-(\ref{grrpq}), (\ref{cubeq})-(\ref{Rirz}),  (\ref{pqrz}), (\ref{Spq}) and (\ref{transfT1}).
One can easily obtain the transformation inverse to (\ref{transfT1}) in the form
\BEA
t&=&\pm\sqrt{\sqrt{\br^2+(\bz-\bz_3)^2}-(\bz-\bz_3)}\sinh \bt  \ , \nonumber \\
\rho&=&\sqrt{\sqrt{\br^2+(\bz-\bz_3)^2}+(\bz-\bz_3)} \ , \label{transfT1i} \\
z&=&\pm\sqrt{\sqrt{\br^2+(\bz-\bz_3)^2}-(\bz-\bz_3)}\cosh \bt \ , \nonumber
\EEA
where either upper signs or lower signs are valid.

The form (\ref{BStvar}) represents a general boost-rotation symmetric spacetime
``with rotation", i.e.,  with the Killing vectors which are not hypersurface orthogonal.
Putting $A=0$, we recover the canonical form of the boost-rotation symmetric spacetimes with the
hypersurface orthogonal Killing vectors the general structure of which was studied in
\cite{BicSch} . Notice that transformation (\ref{transfT1}) is meaningful only for $z^2 > t^2$,
i.e.,  ``below  the roof",  using the terminology of \cite{BicSch}.
Here it leads to the explicit  forms of metric functions  ${\rm e}^{\mu}$,  ${\rm e}^{\lambda}$,
and $A$.  Assuming then that these functions are analytically continued across the roof
$z^2=t^2$ and  across the null cone  $z^2+\rho^2-t^2=0$, we find out that,  by choosing the coefficients $\kappa_1$,  $\kappa_2$, $\kappa_3$, $\kappa_4$  in the relation
(\ref{eqtrKill}) appropriately, the metric
can be made smooth across both the roof and  null cone except for the axis $\rho=0$ in general,
 as will be seen in the following. (In the original coordinates $\{\tau,p,q, \sigma\}$ the region $t^2>z^2$
corresponds to the square $\{3,3\}$ in Fig. 2.)
In addition, as a consequence of transformation (\ref{transfT1}), (\ref{transfT1i}), we now get {\it two} rotating
black holes uniformly accelerated in opposite directions, instead of one represented in the original
Weyl coordinates,  since the transformation (\ref{transfT1i}) with the upper signs
transforms the Weyl region into the region with $z^2>t^2$, $z>0$,
whereas that with the lower signs into $z^2>t^2$, $z<0$
(see Fig. 4  and discussion below). 

In order to choose the coordinates $\bt$, $\bfi$, $t$, $\varphi$ and the Killing vectors appropriately, i.e.,  to fix constants
$\kappa_1 \dots \kappa_4$ in the relation (\ref{eqtrKill}), we analyze the behaviour
of the metric at spacelike infinity. This is simpler  to do in the original coordinates $p$, $q$ where 
we look for a curve approaching 
\BE
\rho \rightarrow \infty \ , \ \  \varphi \rightarrow {\rm{const}}, \ \
z \rightarrow {\rm{const}}  \ , \ \  t \rightarrow {\rm{const}} \ , \ \  z^2>t^2 \ .
\label{krivka}
\EE
In Weyl's coordinates this can be achieved by choosing a parameter  $v$ and putting $\br=v$, $\bz=v^2$, $\bt=$const, $\bfi=$const,
with $v \rightarrow \infty$. In the coordinates $p$, $q$ we can choose, correspondingly, the curve
\BE
p=p_2+v^{-2} \ , \quad q=-p_2+v^{-4} \ ,  \label{krivkapq}
\EE
$p_2$ being the root of $P$, with $v \rightarrow \infty$. This  curve approaches the point $L$ in Fig. 3.
Regarding the transformation (\ref{tranz}), (\ref{tranr}), we see that  indeed  $\br \sim v$, $\bz \sim v^{2}$,
$\bt =$const, $\bfi=$const, as required. 

In order to obtain spacetimes which are asymptotically Minkowskian  along the curve (\ref{krivka}),
we now require
\BE
{\rm e}^{\mu} \rightarrow 1 \ , \quad {\rm e}^{\lambda} \rightarrow 1 \ , \quad A \rightarrow 0
\label{asymp}
\EE
for (\ref{krivka}), i.e., along the curve (\ref{krivkapq}) with $v \rightarrow \infty$. The explicit
form of $g_{\br \br}$, given  in (\ref{grrpq}), when expressed along the curve (\ref{krivkapq}),
leads to $g_{\br \br} = {\rm e}^{2 \nu}/{\rm e}^{2 U} \sim v^{-2} \sim \br^{-2}$.
The requirement (\ref{asymp}) rewritten in terms of functions $U$ and $\nu$,
leads to
 ${\rm e}^{2 U} \rightarrow$ const, ${\rm e}^{2 \nu} \sim  \br^{-2} \sim \rho^{-2} $
(cf. Eqs.  (\ref{transfT1}), (\ref{trel})); it  is thus satisfied if  ${\rm e}^{2 U} \rightarrow$ const.
This implies, using Eqs. (\ref{e2Upq}) and (\ref{gCcompts}), the following relation for the coefficients 
$\kappa_3$ and $\kappa_1$:
\BE
\kappa_3 = -p_2^2 \kappa_1.
\EE
Similarly, requiring $A \rightarrow 0$ along the curve ({\ref{krivka}}), we deduce
 from Eqs. (\ref{apq}) and (\ref{gCcompts}) the relation for $\kappa_2$ and $\kappa_4$:
\BE
\kappa_2 = p_2^2 \kappa_4.
\EE

Until now we considered the region $z^2>t^2$,  i.e., ``bellow the roof".  Turning  to the roof
itself, the necessary  condition of the regularity of the metric  there  reads
(cf. the detailed discussion in Section V in \cite{BicSch} in the case of the hypersurface
orthogonal Killing vectors) 
\BE
{\rm e}^{\lambda} {\rm e}^{-\mu} = 1 \quad {\rm at} \quad  z^2=t^2\ .
\EE
Eq.   (\ref{trel}) then implies
\BE
\frac{{\rm e}^{2 \nu}}{{\rm e}^{4 U}} \br^2  \rightarrow 1 \  \ {\rm for} \ \  \br \rightarrow 0\ , \ z > z_3 \ , 
\label{podmas}
\EE
or, in coordinates $p$, $q$,  in the limit when $q \rightarrow -p_2=+q_3$, $p \in  \langle p_2, p_3 \rangle$,
so that the lower segment $L\bz_3$  in Fig. 3  is approached. 
Calculating this limit we find that the condition  (\ref{podmas})  is satisfied if the coefficient
$\kappa_1$ is given by 
\BE
\kappa_1^2 = \left(  p_2 (m^2 p_2^3 - 2 \e m p_2^2  +(4 \g^2 +\e^2) p_2 -4m\g )\right)^{-1} \ .
\label{condA1}
\EE

Finally we wish to study the regularity of the axis  $\rho=0$. Clearly, with the SC-metric
whole axis cannot be regular. There will be black-hole horizons mapped on  the segments
of the axis, along other parts string singularities  will in general be located. These can
be understood as the ``cause" of the acceleration  of the holes. However, pieces of the axis
can be made regular (see again \cite{BicSch} for a detailed discussion of the regularity
of the axis of general boost-rotation symmetric spacetimes with hypersurface orthogonal
Killing vecors). 
The regularity of the axis  requires
\BE
\lim_{\rho_0 \to 0}
\frac{1}{2 \pi} \frac{\int_0^{2 \pi} \sqrt{g_{\varphi \varphi}}|_{\rho_0} {\rm d} \varphi}{\int_0^{ \rho_0} \sqrt{g_{\rho \rho}} {\rm d} \rho} = 1 \label{regInt} \  ,
\EE
which implies the condition 
\BE
{\rm e}^{\lambda+\mu} \rightarrow 1 \quad  {\rm at}  \quad \rho=0 \ . \label{regosaBS}
\EE
This is easy to see since  from Eq. (\ref{BStvar}) we have 
\BE
g_{\rho \rho} = {\rm e}^{\lambda} \ , \ g_{\varphi \varphi} = \rho^2 {\rm e}^{-\mu} - A^2 {\rm e}^{\mu} (z^2-t^2),
\EE
but calculating $\lim_{\rho \to 0} A$ from Eqs. (\ref{apq}), (\ref{gCcompts}),  (\ref{pqrz}) and
(\ref{transfT1}),  we find 
\BE
A=O(\rho^2) \quad  {\rm as} \ \rho \rightarrow 0 \quad {\rm for} \quad z > Z_1\ , \ \  z < - Z_1 \ ,
\label{Aout}
\EE
where
\BE
Z_1 = \left( 2(\bz_3-\bz_1) + t^2 \right)^{1/2} \ .  \label{eqZ1}
\EE
Hence, the condition (\ref{regosaBS})  requires ${\rm e}^{2 \nu} \rightarrow 1 \ \ {\rm for } \ \ p \rightarrow p_2 \ , \  -p_2 < q < -p_1 $ (cf. Fig. 3). This  implies 
\BE
 {\rm e}^{\nu} \ \rightarrow \kappa_4^{-2} \kappa_1^{2} \ ,
\EE
so that the axis will be regular in the regions extending from the holes to infinity if
\BE
\kappa_4^2 = \kappa_1^2.
\EE

Therefore, we have now fixed all constant parameters $\kappa_1 \dots \kappa_4$ in terms of the 
original parameters $m$, $\e$ and $\g$ (the root $p_2$ being determined by the solution of $P=0$
where $P$ is given by Eq. (\ref{eqP})).
The axis is thus regular for $z^2>Z_1$. As a consequence  of transformation (\ref{transfT1}) we find
two accelerated black holes ``located" at the axis at the two segments $\langle Z_2, Z_1\rangle$
and $\langle -Z_1, -Z_2\rangle$, where $Z_1$ is given by expression (\ref{eqZ1}) and
\BE
Z_2 = \left( 2(\bz_3-\bz_2) + t^2 \right)^{1/2} \ .
\EE
The situation is illustrated in Fig. 4a.
As in the Weyl picture (and in the case of a non-rotating C-metric) we do not get the ``inner parts"
of the holes - these can be obtained only by analytic continuations across the horizons located
at  $\langle Z_2,  Z_1 \rangle$ and $\langle -Z_1, -Z_2 \rangle$. 
However, it is not difficult to see that the Killing vector
$\partial/\partial \bt$ becomes null outside the axis - this corresponds to the outer boundary of the
ergoregions. 

\begin{figure}
\begin{center}
\includegraphics*[height=5cm]{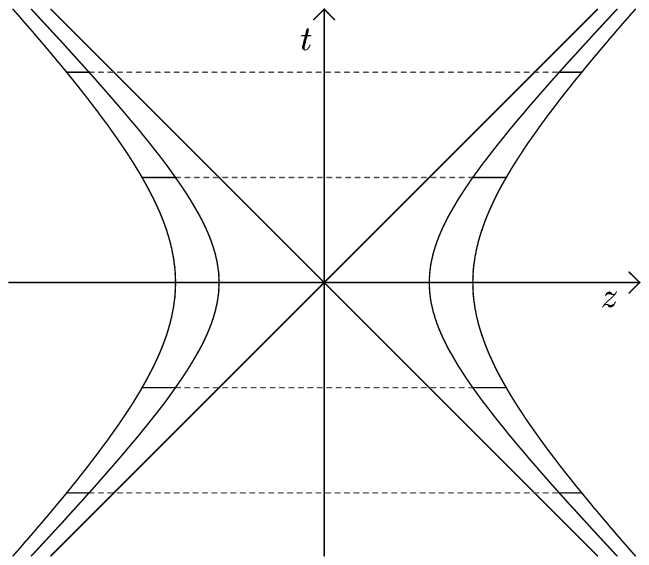}
\hspace{2cm}
\includegraphics*[height=5cm]{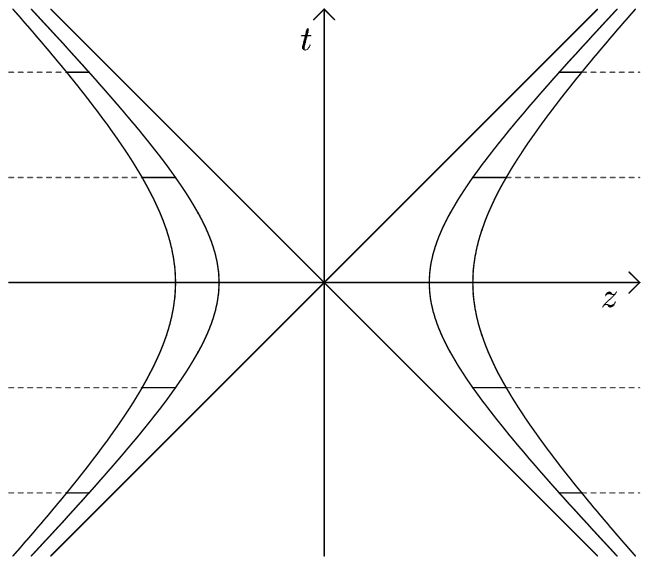}
\end{center}
\caption{Two uniformly accelerated spinning black holes\\
   a) connected by a spring between them, \\
   b) with string extending from each of them to infinity.}
\end{figure}

Between the accelerated holes, i.e.,  for $z \in \langle -Z_2,Z_2 \rangle$,  a nodal (string)  singularity
occurs (violating conditions (\ref{regInt}), (\ref{regosaBS})) which can be understood
as causing the  accelerations of the holes away from each other. 
It is interesting to observe that there exist
causality violation regions around this nodal singularity (here $g_{\varphi \varphi} <0$)  which tend to
be dragged along with black holes. A detailed discussion  of these regions, as well as of the
ergoregions of the SC-metric, will be given elsewhere.

Here, let us yet notice that, similarly  to the non-rotating  C-metric and, indeed, to the case of all boost-rotation symmetric
spacetimes \cite{BicSch},  we can construct SC-metrics in which the axis is regular between the holes
and strings are located outside them, i.e.,  at $z>Z_1$,  $z<-Z_1$  (see Fig 4 b). As a
consequence of the field equations function $A$ in the metric (\ref{BStvar}) is constant along the axis outside
the horizon, its value being different at $z>Z_1$,  $z<-Z_1$ from that  at  $z \in \langle -Z_2, Z_2 \rangle$.
(This can be  checked by analyzing directly the expression for $A$ given  by Eqs. (\ref{apq}), (\ref{gCcompts}),
(\ref{pqrz}),(\ref{Spq}), (\ref{transfT1})). 
In the above case when the nodal singularity extends between the horizons, $A=0$ at the axis for
$z>Z_1$, $z<-Z_1$ (see Eq. (\ref{Aout})), but $A=A_0 \not= 0$ at $\rho=0$, $z \in \langle -Z_2, Z_2 \rangle$.
Since field equations allow us to add an additive constant to $A$, we may just take $\tilde A (\rho,z,t)
=  A (\rho,z,t)-A_0$ which implies a regular axis between the holes but nodal singularities at
$z^2>Z_1^2 $ (see Fig. 4 b).

\section{Concluding remarks}
Our main result is the explicit representation (\ref{BStvar})-(\ref{trel}) of the spinning
C-metric as a spacetime with an axial and a boost Killing vector which are not hypersurface 
orthogonal. This has been achieved by first transforming the original form of the metric
into the Weyl coordinates and, subsequently, by going over to coordinates adapted
to boost-rotation symmetry. Simultaneously, two new ``radiative" regions of the spacetime
arise in which the boost Killing vector is spacelike.

Indeed, it is easy to see that under the transformation (\ref{transfT1}), (\ref{transfT1i}),
the Killing vector $\xi_{(\bt)}= \partial/\partial \bt$ 
goes over into
\BE
         \xi = z \frac{\partial}{\partial t} + t \frac{\partial}{\partial z}  \ .                              
\EE
This is the {\it boost} Killing vector which in the coordinates
adapted to the boost-rotation symmetry of the SC-metric  has everywhere
the same form as in flat space. Its norm is given by
\BE
            |\xi|^2 = - {\rm e}^{\mu}  (z^2-t^2) \ .                     \label{normboost}
\EE
One can make sure that, after analytically continuing the metric (\ref{BStvar})
into the region $t^2>z^2$, the norm (\ref{normboost}) is always positive there so that
the Killing vector is spacelike.

Very recently Letelier and Oliveira  \cite{Letelier}  gave an incomplete transformation of the SC-metric into the
Weyl form.
However, these authors did not choose  the Killing vectors and, thus, also the corresponding
coordinates $\bt$ and $\bfi$ which would reflect the boost-rotational symmetry of the SC-spacetimes.
In fact in contrast to our Eq.  (\ref{eqtrKill}) relating ($\tau$, $\sigma$) to ($\bt,\bfi$), they put simply
$\tau \sim \bt$, $\sigma \sim \bfi$. In addition, their formula for the radial coordinate is not correct
(in their last Eq. (34) the second term proportional to $(a/m)^2$ should not appear). This leads
to an incorrect ``location" of the axis in their Fig. 1 etc.

From the form  (\ref{BStvar}) we have seen (cf. Eq. (\ref{asymp})) that the coordinates can 
be chosen so that the metric becomes  explicitly Minkowskian at spatial infinity. We did not
analyze rigorously the structure of null infinity where we expect to find a non-vanishing
radiation field as it exists in the standard C-metric and, indeed, in all boost-rotation
symmetric spacetimes \cite{BicPra}. Here we confine ourselves to giving Fig. 5
which demonstrates  the radiative character of the SC-metric. In the figure
the  invariant proportional to $I_3^{\frac{1}{12}}$ and $I_3^{\frac{1}{6}}$, where
$I_3$ is given by Eq. (\ref{invars}), is plotted at a fixed time $t=t_0>0$ and at different
times. The radiative field has the character of a pulse, as  was noticed
many years ago in \cite{Bic68} for the special  boost-rotation symmetric solutions
with hypersurface orthogonal Killing vectors. It is also of interest to plot the gradient
of $g_{\varphi t} \sim A$ as a function of time: here again the pulse character can be observed.

Although we did not here analyze the radiative properties of  the SC-metric rigorously,
it is most plausible to expect that in the situation described by Fig. 4b one can give
arbitrarily strong hyperboloidal data (see \cite{Friedrich} for a recent review  and references on this 
approach to the initial value problem)
 on a hyperboloid ``above the roof" $(t>|z|)$, which
represent the SC-metric data and which evolve into the spacetime with smooth null infinity
with a non-vanishing  radiation field. No other spacetime of this type, with two Killing
vectors which are not hypersurface orthogonal, is available in an explicit form. This 
suggests the SC-metric to become  a useful, non-trivial test-bed for numerical relativity. 

\vspace{2cm}

\begin{figure}
\begin{center}
\includegraphics*[height=7cm]{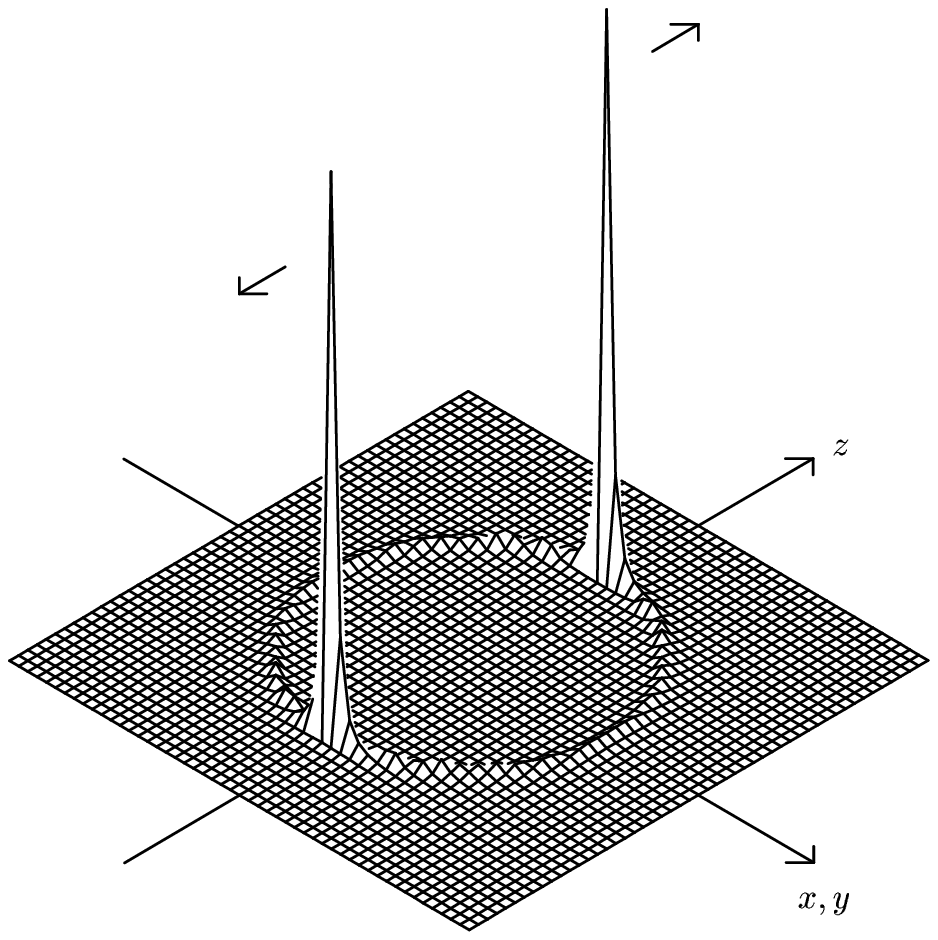}
\hspace{1cm}
\includegraphics*[height=5cm]{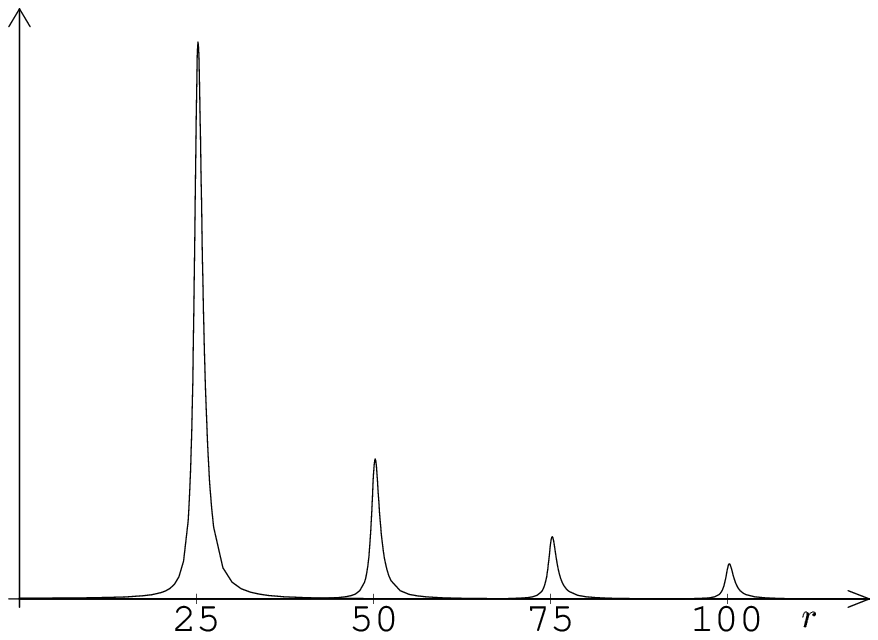}
\end{center}
\caption{The plot of the curvature invariants related to $I_3$ in (\ref{invars}) at fixed $t=t_0>0$ (Fig. 5a). The peaks are located ``above" the accelerated, 
rotating black holes. The gravitational radiation pulse  propagates in all directions  with velocity of light
and decreasing amplitude as seen from Figure 5b, plotted for $\theta=\pi/2.$}
\end{figure}

{{\bf{Acknowledgments}}}\\[2mm]
V. P. enjoyed the hospitality of the Institute of Theoretical Physics of  the F. Schiller-University,
Jena, where part of this work was done. Helpful discussions with Alena Pravdov\'a are gratefuly
acknowledged. J. B. thanks Swiss National fonds for support and Petr H\'aj\'\i \v cek for discussions
and kind hospitality at the Institute of Theoretical Physics in Berne, where this work was finished.
We were supported in part by the Grant  No. GACR-209/99/0261 and V. P.
also by the Grant No. GACR-201/98/1452.


\begin{thebibliography}{99}
\bibitem{Kramer}
D. Kramer, H. Stephani, M. MacCallum and E. Herlt, {\it Exact solutions of  Einstein's   Field Equations},
(Cambridge University Press, 1980).
\bibitem{KW}
W. Kinnersley and M. Walker,  {\it Uniformly accelerating charged mass in General Relativity},
Phys. Rev. D {\bf 2}, 1359 (1970).
\bibitem{Ernst1}
F. J. Ernst, {\it Generalized C-metric}, J. Math. Phys. {\bf 19}, 1986 (1978);  W. B. Bonnor, 
{\it  An exact solution of Einstein's equations for two
particles falling freely in an external gravitational field},
Gen. Rel. Grav. {\bf 20},
 607 (1988).
\bibitem{Ernst2}
F. J. Ernst,  {\it  Removal of the nodal singularity of the C-metric},  J. Math. Phys.  {\bf 17}, 515  (1976).
\bibitem{Hawking1}
S. W. Hawking, G. T. Horowitz and S. F. Ross,  {\it  Entropy, area, and black hole pairs}, Phys. Rev. D {\bf 51}, 4302 (1995).
\bibitem{Hawking2}
S. W. Hawking and  S. F. Ross, {\it Pair production of black holes on cosmic strings},  Phys. Rev. Lett.  {\bf 75}, 3382 (1995).
\bibitem{Mann}
R. B. Mann and  S. F. Ross, {\it  Cosmological production of charged black hole pairs},  Phys. Rev. D {\bf 52}, 2254 (1995).
\bibitem{AshD}
A. Ashtekar and T. Dray, {\it On the existence of solutions to Einstein's equation with non-zero
Bondi-news},  Comm. Phys. {\bf 79}, 581 (1981).
\bibitem{Bonnor}
W. B. Bonnor,  {\it The sources of the vacuum C-metric},  Gen. Rel. Grav. {\bf 15}, 535 (1983).
\bibitem{Cornish1}
F. H. J. Cornish and W. J. Uttley, {\it  The interpretation of the C metric. The Vacuum case},  Gen. Rel. Grav. {\bf 27}, 439 (1995).
\bibitem{Cornish2}
F. H. J. Cornish and W. J. Uttley, {\it  The interpretation of the C metric. The charged case when
$e^2 \leq m^2$},  Gen. Rel. Grav. {\bf 27}, 735 (1995).
\bibitem{Wells}
C. G. Wells, {\it Extending the black hole uniqueness theorems I.},  gr-qc/9808044, (1998).
\bibitem{BicSch}
J. Bi\v c\' ak and  B. G. Schmidt, {\it Asymptotically flat radiative space-times with boost-rotation
symmetry: The general structure}, Phys. Rev. D {\bf 40}, 1827 (1989).
\bibitem{Bic68}
J. Bi\v c\'ak, {\it Gravitational radiation from uniformly accelerated particles in general relativity},
 Proc. Roy. Soc.  A {\bf 302}, 201 (1968).
\bibitem{BicT}
J. Bi\v c\'ak,  {\it  Radiative properties of space-times with the axial and boost symmetries}, in {\it Gravitation and Geometry}, edited by
W. Rindler and A. Trautman, (Bibliopolis, Naples, 1987).
\bibitem{BicB}
J. Bi\v c\'ak, {\it On exact radiative solutions representing finite sources}, 
 in {\it Galaxies, Relativity and Axisymmetric  Systems and Relativity, Essays presented
 to W. B. Bonnor on his 65th birthday}, edited by M. A. H.  MacCallum, (Cambridge University Press,
Cambridge, 1985).
\bibitem{PP}
V. Pravda and A. Pravdov\' a, {\it Uniformly accelerated sources in electromagnetism and
 gravity}, in {\it Proceedings of the week of postgraduate students}, (Charles university, Prague 1998);  gr-qc/9806114 (1998).
\bibitem{LesHouches}
J. Bi\v c\'ak, {\it Radiative spacetimes: exact approaches}, in {\it Relativistic Gravitation and Gravitational Radiation, Proceedings of the Les Houches
School of  Physics},  edited by J. A. Marc and  J. P. Lasota, (Cambridge University Press,
Cambridge,  1995).
\bibitem{BicPra}
J. Bi\v c\'ak, A.  Pravdov\'a, {\it  Symmetries of asymptotically flat electrovacuum spacetimes and radiation},
J. Math. Phys. {\bf 39}, 6011  (1998).
\bibitem{Pleb}
 J. F. Pleba\'nski and M. Demia\'nski, {\it Rotating, charged and uniformly accelerating
 mass in general relativity},  Annals of Phys. {\bf 98},  98   (1976).
\bibitem{FZ}
H. Farhoosh and R. L. Zimmerman,{\it Surfaces of infinite red-shift around a uniformly accelerating
and rotating particle}, Phys. Rev. D {\bf 21}, 2064 (1980).
\bibitem{Letelier}
P. S. Letelier and S. R. Oliveira, {\it On Uniformly Accelerated Black Holes},  gr-qc/9809089 (1998).
\bibitem{Bonnor2}
W. B. Bonnor, {\it The C-metric with $m=0,\ e \not=0$}, Gen. Rel. Grav. {\bf 16}, 269  (1984).
\bibitem{Friedrich}
H. Friedrich, {\it Einstein's Equations and Conformal Structure},
     in {\it The Geometric Universe. Science, Geometry, and the Work of Roger
     Penrose}, eds. S. A. Huggett et al, Oxford Univesrity Press, pp. 81-97,  (1998).
\end{thebibliography}
 \end{document}